\documentclass[a4paper,10pt]{article}

\usepackage{setspace}

\usepackage[usenames]{color}
\usepackage[table]{xcolor}
\usepackage{soul}
\usepackage{textcomp}
\usepackage{multicol}  
\usepackage{multirow} 
\usepackage{xspace}
\usepackage{amsmath} 
\usepackage{amssymb}
\usepackage{graphicx}
\usepackage{breqn}
 \usepackage[bookmarks=false]{hyperref}
\usepackage{algorithm}
\usepackage{algorithmic}
 \usepackage{cuted}
 \hypersetup{bookmarks={false}}
  
\newtheorem{definition}{Definition}

\newcommand*{\QEDB}{\hfill\ensuremath{\square}}%

\def\V{\mbox{$\mathcal{V}$}} %

\newcommand{\mpara}[1]{\medskip\noindent{\bf #1}}

\def\fftwyt{FF-TW-YT}
\def\ghsotw{GH-SO-TW}

\begin{document}

\title{Revisiting Resolution  and Inter-Layer Coupling  Factors in Modularity for Multilayer Networks}

\author{Alessia Amelio, Andrea Tagarelli 
\\DIMES, University of Calabria\\
87036 Arcavacata di Rende (CS) - Italy\\
Email: \{aamelio,tagarelli\}@dimes.unical.it 
}

\date{August 5, 2017}


%

\maketitle
\begin{abstract}
Modularity for multilayer networks, also called multislice modularity, is parametric to a resolution factor and an inter-layer coupling factor.  The former is useful to express layer-specific relevance and the latter quantifies the strength of  node linkage across the layers of a network.  However,  such parameters can be set arbitrarily, thus discarding  any structure information at graph or community level.  Other issues are related to the inability of properly  modeling order relations over the layers, which is required for dynamic networks.  
 
 In this paper we propose a new definition of modularity for multilayer networks that aims to overcome major issues of existing multislice modularity.  We revise the role and semantics of   the layer-specific resolution   and   inter-layer coupling terms, and define parameter-free unsupervised approaches for their computation, by using  information from the within-layer and inter-layer structures of the communities.  Moreover, our formulation of multilayer modularity is general enough to account for an available ordering of the layers and relating constraints on layer coupling. Experimental evaluation was conducted using three  state-of-the-art methods for multilayer community detection and nine real-world multilayer networks. Results have  shown the significance of our   modularity,  disclosing  the  effects of different combinations of the resolution and inter-layer coupling functions.  This work can pave the way for the development of new optimization methods for  discovering community structures in multilayer networks.
\end{abstract}


\section{Introduction}
 
The concept of \textit{modularity}  originally introduced by Newman  and Girvan in~\cite{Newman04} is a well-known quality criterion for graph clustering problems. It has  been widely used in optimization methods for discovering   \textit{community structure} in networks~\cite{BrandesDGGHNW08,CarchioloLMM10,ChenKS14}, through 
greedy agglomeration~\cite{Newman04b,Clauset04},  spectral division~\cite{Newman05,WhiteS05}, simulated annealing~\cite{Reichardt06}, 
  or extremal optimization~\cite{Duch05},  just  to mention a few prominent, follow-up studies.  Since then, there has been a continuously growing interest in this measure, which   has prompted a variety of computational frameworks (see, e.g.,~\cite{2010Tang} for a comprehensive book on this matter).
   
The key assumption behind modularity is that a network  can be organized into modules, a.k.a. \textit{communities} or \textit{clusters}, such that there are fewer edges than expected (rather than few edges between communities as determined by cut-ratio). 
Therefore, modularity quantifies the difference between the  expected  and the  actual   number of edges linking nodes inside a community. 
This expectancy is expressed through a configuration graph model, having a certain degree distribution and randomized edges. Since this graph ignores any community structure, a large difference between actual connectivity and expected connectivity would   indicate 
 community structure.  

Modularity has been also extended to the   general case of \textit{multilayer networks}. Multilayer networks   
are increasingly used as a powerful model to represent the organization and relationships of complex data  in a wide range of scenarios~\cite{Magnanibook,Kivela+14}. In particular, a great deal of attention has recently been devoted to the problem of  community detection in a multilayer network~\cite{Mucha10,KimL15}. 
Solving this problem is  important in order to unveil meaningful patterns of node groupings into communities,  by taking into account the different interaction types that involve all the entity nodes in a complex network.    
    Mucha et al.~\cite{Mucha10}  presented a general framework to study the community structure of arbitrary multilayer networks (also called \textit{multislice} in that work), by extending the modularity function to those  networks.  
 
 Mucha et al.'s modularity involves two additional parameters w.r.t. classic modularity: a \textit{resolution} parameter  and an \textit{inter-layer coupling} factor. The former, by modeling a notion of layer-specific relevance, allows for   controlling the effect on the size distribution of community due to the \textit{resolution limit} known in modularity~\cite{FortunatoB07}. The latter factor,   instead, quantifies the strength of  linking of nodes across different layers.  
 However, there are  a few limitations and issues in this measure. In particular, the resolution parameter can be  arbitrarily   set for each layer, regardless of any structure information at graph or community level. Moreover, the inter-layer coupling term weights the involved layers in a  uniform way. Yet,  all pairs of layers can in principle be considered for the coupling parameter, which  makes no sense in certain scenarios such as modeling of time-evolving networks. 
 
Motivated by the above remarks, in this work we propose a new modularity measure for multilayer networks, which aims to overcome all of the aforementioned issues of the earlier    measure. 
By using  information from the within-layer and inter-layer structures of the communities, 
we define the resolution factor in function of   each community and layer, and  different  inter-layer coupling schemes that account  for properties of a community projection over any two  comparable  layers.  
More specifically, our proposed resolution term   exploits the notion of \textit{redundancy} to account  for the strength of the contribution that   a layer takes  in determining the internal connectivity for each community. We define different approaches for measuring the inter-layer coupling term, which is based on the relevance of the projection of a community w.r.t. two coupled layers.  
Our formulation of multilayer modularity is general enough to account for an available ordering of the layers, and relating constraints on their coupling validity.

Experimental evaluation was  conducted mainly to understand   how the proposed multilayer modularity behaves w.r.t. different settings regarding the resolution   and the inter-layer coupling terms.  Using   state-of-the-art methods for multilayer community detection (GL, PMM, and LART) and 9 real-world multilayer networks,  interesting remarks have arisen from results,   highlighting  the significance of our formulation and the different  expressiveness against the previously existing multislice    modularity.


\section{Background} 
\label{sec:background}

\subsection{Modularity}

Given an undirected graph $G=(\V,\mathcal{E})$, with $n=|\V|$ nodes and $m=|\mathcal{E}|$ edges,  let $\mathcal{C}$ be a community structure over $G$. For any  $v \in \V$,  we use   $d(v)$ to denote the degree of $v$, and for any community $C \in \mathcal{C}$, symbol $d(C)$ to denote the degree of   $C$, i.e.,  $d(C)=\sum_{v \in C} d(v)$; also,  the  total degree of nodes over the entire graph,     $d(\V)$, is defined as $d(\V)=\sum_{v \in \mathcal{V}} d(v)=\sum_{C \in \mathcal{C}} d(C)=2m$.  
Moreover, we denote with $d_{int}(C)$  the internal degree of $C$, i.e., the portion of $d(C)$ that corresponds to the number of edges linking nodes in $C$ to other nodes in $C$. 
Newman and Girvan's \textit{modularity} is defined as follows~\cite{Newman04}:
\begin{equation}\label{eq:NGmodularity}
Q_{NG}(\mathcal{C})=  \sum_{C \in \mathcal{C}} \frac{d_{int}(C)}{d(\V)} - \left( \frac{d(C)}{d(\V)} \right)^2  
\end{equation}
 
 In the above equation, the first term is maximized when many edges are  contained in clusters, whereas the second term is minimized   by partitioning  the graph into many clusters with small total degrees. 
The value of $Q_{NG}$ ranges within  -0.5 and 1.0~\cite{BrandesDGGHNW08} --- it is  minimum  when all edges link nodes in different communities, and  maximum   when all edges link nodes in the same community. 

\subsection{Multilayer network model}
Let $G_{\mathcal{L}} = (V_{\mathcal{L}}, E_{\mathcal{L}}, \V, \mathcal{L})$ be a \textit{multilayer network} graph, where $\V$ is a set of entities  and  $\mathcal{L}= \{L_1, \ldots, L_{\ell}\}$ is a set of layers. Each layer represents a specific type of relation between entity nodes. 
%
Let $V_{\mathcal{L}} \subseteq \V \times \mathcal{L}$ be the set containing the entity-layer combinations, i.e., the occurrences of each   entity  in the   layers.   $E_{\mathcal{L}} \subseteq V_{\mathcal{L}} \times V_{\mathcal{L}}$ is the set of undirected links connecting the entity-layer elements. 
 For every  $L_i \in \mathcal{L}$, we define $V_i = \{v \in \V \ | \    (v, L_i) \in V_{\mathcal{L}}\} \subseteq \mathcal{V}$ as the set of nodes in the graph of $L_i$, and $E_i \subseteq V_i \times V_i$ as the set of edges in $L_i$. Each entity must be present in at least one layer, i.e., $\bigcup_{i = 1..\ell} V_i  = \V$, but each  layer is not required to contain all  elements of $\V$. 
We assume that the inter-layer links only connect the same entity in different layers, however  each entity in one layer could be  linked to the same entity in a few or all other layers.

\subsection{Multislice Modularity}
\label{sec:Qms}
Given a community structure   $\mathcal{C}$ identified over a multilayer network $G_{\mathcal{L}}$, the \textit{multislice modularity}~\cite{Mucha10} of   $\mathcal{C}$ is defined as:
\begin{eqnarray}\label{muchaMod}
Q_{\textrm{ms}}(\mathcal{C})\!\!\!\!\!&\!\!=\!\!&\!\!\!\!\frac{1}{d(V_{\mathcal{L}})}  \sum_{\substack{u,v, \\ L_i,L_j}}   \left[  \left( A_{uvL_i} - \gamma_i \frac{d_{L_i}(u)d_{L_i}(v)}{2E_i} \right)  \delta_{L_i,L_j} \right. + \nonumber \\
  & + &  \left. \delta_{u,v}\mathrm{C}_{v,L_i,L_j} \right] \delta(g_{u,L_i}, g_{v,L_j})
\end{eqnarray}
where 
$d(V_{\mathcal{L}})$ is the total degree of the multilayer network graph, 
$d_{L_i}(u)$ denotes the degree of node $u$ in layer $L_i$, 
$A_{uvL_i}$ denotes a link between $u$ and $v$ in $L_i$,  
 $2E_i$ is the total degree of the graph of layer $L_i$, $\gamma_i$ is the resolution parameter for layer $L_i$,   $\mathrm{C}_{v,L_i,L_j}$ quantifies the links of node $v$ across  layers $L_i$,  $L_j$.  
Moreover, the Dirichlet terms have the following meanings: 
$\delta_{L_i,L_j}$ is equal to 1 if $L_i=L_j$ and 0 otherwise, 
$\delta_{u,v}$ is equal to 1 if $u=v$ and 0 otherwise (i.e., the inter-layer coupling is allowed only for nodes corresponding to the same entity), 
and 
$\delta(g_{u,L_i}, g_{v,L_j})$ is equal to 1 if the community assignments of node $u$ in $L_i$ and node $v$ in $L_j$ are the same and 0 otherwise.  



\mpara{Limitations of $Q_{\textrm{ms}}$.\ \ }
As   mentioned in the Introduction, a different   resolution parameter $\gamma$ can be associated with each layer to express the weight of its relevance; however, the authors do not clearly specify any principled way to set a layer-weighting scheme. 
More critically, neither the inter-layer coupling term  (i.e., $C_{v,L_i,L_j}$) or any constraint on the layer comparability are clearly  defined; actually, the authors simply choose to set all nonzero inter-layer edges to a constant value $\omega$, for all unordered pairs of layers.    
Yet, both $\gamma_i$  and $\omega$ parameters can   assume any non-negative value, which further increases a clarity issue in the     properties of $Q_{\textrm{ms}}$. 
    

\section{Proposed Multilayer Modularity}
In this section, we propose a new definition  of   modularity for multilayer networks that aims to overcome all of the issues of $Q_{\textrm{ms}}$ previously discussed. 

Our goal is to  reconsider the role and semantics of the two key elements in multilayer modularity: the \textit{layer-specific resolution}   and the \textit{inter-layer coupling}.  
For both terms, we provide formal definitions, which avoid any user-specified setting possibly based on a-priori assumptions on the network.    By contrast, we  conceive  parameter-free unsupervised approaches for their computation, by using  information from the within-layer and inter-layer structures of the communities.  
More specifically, we   define the  resolution factor in function of any given pair of layer and community, and define the inter-layer coupling term to account for properties of community projection over any two comparable layers. 
%
Moreover,  we  
 consider the introduction of a \textit{partial order relation} $\prec_{\mathcal{L}}$ over the layers  in order to properly represent  scenarios in which a particular ordering among layers is required. 
 To distinguish from the conventional case of $\mathcal{L}$ as an unordered set, we might refer to  notation $G_{\mathcal{L}}^{\prec}=(G_{\mathcal{L}}, \prec_{\mathcal{L}})$, which couples $G_{\mathcal{L}}$ with the order relation $\prec_{\mathcal{L}}$. 
%

 \vspace{2mm}
\begin{definition}[Multilayer Modularity]
\em
Let $G_{\mathcal{L}} = (V_{\mathcal{L}}, E_{\mathcal{L}}, \V, \mathcal{L})$ be  a multilayer network graph and, optionally, let  $\prec_{\mathcal{L}}$ be  a partial order relation over the set of layers $\mathcal{L}$. 
Given a community structure $\mathcal{C}=\{C_1, \ldots, C_k\}$ over $G_{\mathcal{L}}$, the   \emph{multilayer modularity}  is defined as: 
\end{definition}
 %
%
\begin{equation}\label{eq:MLmodularity}
Q(\mathcal{C}) =  \frac{1}{d(V_{\mathcal{L}})}\sum_{C \in \mathcal{C}} \left[ \sum_{L \in \mathcal{L}}  \left(  d_L^{int}(C)  - \gamma(L,C) \frac{(d_L(C))^2}{d(V_{\mathcal{L}})}   + \beta  \sum_{L' \in \mathcal{P}(L)}  IC(C, L, L')  \right) \right] 
\end{equation} 
%
{\em 
where for any $C \in \mathcal{C}$ and $L \in \mathcal{L}$:
\begin{itemize}
\item   $d_L(C)$ and $d_L^{int}(C)$ are  the degree of  $C$ and the internal degree of  $C$, respectively, by considering only edges  of layer $L$; 
\item   $\gamma(L,C)$ is the value of the  resolution function;  
 \item   $IC(C, L, L')$ is the value of the inter-layer coupling function for any valid layer pairings with $L$;  
 \item  $\beta \in \{0,1\}$ is a parameter to control the  exclusion/inclusion of inter-layer couplings; 
  and 
\item   $\mathcal{P}(L)$ is  the set of  valid pairings   with   $L$ defined as:
$$
\mathcal{P}(L)=
\begin{cases}
    \{L' \in \mathcal{L} \ | \ L \prec_{\mathcal{L}} L' \}, &  \text{if} \prec_{\mathcal{L}} \text{is defined}\\
    \mathcal{L} \setminus \{L\}, & \text{otherwise}
\end{cases}
$$ 
\end{itemize}
\QEDB
}
 
%

It should be noted that     Eq.~(\ref{eq:MLmodularity}) differs substantially from   Eq.~(\ref{muchaMod}).  Our proposed modularity     originally introduces a resolution factor that varies with each community, and an inter-layer coupling scheme that might also depend on the layer ordering. Moreover,  it  utilizes  the total degree of the multilayer network graph $d(V_{\mathcal{L}})$ instead of the layer-specific degree (i.e., term $2E_{L_i}$, for each $L_i \in \mathcal{L}$). 
 The latter point is important because, as we shall later discuss more in detail,   the total degree of the multilayer graph includes the inter-layer couplings and it might be defined in different ways depending on the scheme of inter-layer coupling.  

In the following, we elaborate on the resolution functional term, $\gamma(\cdot,\cdot)$, and the inter-layer coupling functional term,  $IC(\cdot, \cdot, \cdot)$.

\subsection{Redundancy-based resolution factor} 
The layer-specific resolution factor intuitively expresses the  relevance of a particular layer to the calculation of the expected community connectivity in that layer. 
While this   can always reflect some predetermined scheme of relevance weighting of layers, we propose a more general definition that  accounts for the strength of the contribution that   a layer takes  in determining the internal connectivity for each community.  
The key assumption underlying our approach is   that, since a high quality community should envelope   high information content among its elements, \textit{the resolution   of a layer to control the expected connectivity of    a given community  should be lowered  as its contribution to the information content of the community is higher}. 

In this regard, the \textit{redundancy} measure proposed in~\cite{Berlingerio2011}  is particularly suited to  quantify  the variety of connections, such that it is higher if edges of more types (layers) connect each pair of nodes in the community. 

Let us  denote with $P_1$ the set of node pairs   connected in at least one layer in the graph, and with  $P_2$ the set of ``redundant'' pairs, i.e., the pairs of nodes connected in at least two layers. 
Given a community $C$, $P_1^C$ and $P_2^C$ denote the subset of $P_1$ and the subset of $P_2$, respectively, corresponding to nodes in   $C$. 
The redundancy of $C$, $\rho(C)$,  expresses  the number of pairs in $C$ with redundant connections, divided by  the  number of layers connecting the pairs. Formally:  
\begin{equation}\label{eq:redundancy}
\rho(C) =   \sum_{(v,u) \in P_2^C} \frac{|\{L \in \mathcal{L} \ | \  (v,u,L) \in E_{\mathcal{L}}\}|}{|\mathcal{L}| \times |P_1^C|}  
\end{equation} 
%

Note that in the above formula, the set defined in the numerator of each additive term, refers to the layers on which two nodes in a redundant pair are linked. 
We can hence define the \textit{set of supporting layers} $SL$ for each community $C$ as:
\begin{equation}\label{eq:supportset}
SL(C,\mathcal{L})=\bigcup_{(v,u) \in P_2^C } SL(v,u,\mathcal{L})
\end{equation} 
with $SL(v,u,\mathcal{L})=\{L \in \mathcal{L} \ | \  (v,u,L) \in E_{\mathcal{L}}\}$. 

Using the above defined $SL(C,\mathcal{L})$, we provide the following definition of \textit{redundancy-based resolution factor}.  


\vspace{2mm}
\begin{definition}[Redundancy-based resolution factor]
\em
Given a layer $L$ and a community $C$,   the  \emph{redundancy-based resolution factor} in Eq.~(\ref{eq:MLmodularity}) is defined as:
\begin{equation}\label{eq:redbasedgamma}
\gamma(L,C) = \frac{2}{1+ \log_2(1+ nrp(L,C))}
\end{equation} 
where 
$
nrp(L,C) = |\{ s=SL(v,u,\mathcal{L}) \in SL(C,\mathcal{L})  \ | \ L \in s \}|
$
expresses the number of times  layer $L$ participates in redundant pairs.
\QEDB
\end{definition}

\vspace{2mm}
Note that $\gamma(L,C)$ ranges in $(0, 1]$ as long as $L$ participates in at least one redundant pair, and it decreases as $nrp(L,C)$ increases.  As special case,  it is equal to 2  when $nrp(L,C)~=~0$.

\subsection{Projection-based inter-layer couplings}
We propose a general and versatile approach  to quantify the strength of coupling of nodes in one layer with nodes on another layer. Our key idea is \textit{to determine the fraction of nodes belonging to a community onto a layer that  appears in the projection of the community on another layer, and express the relevance of this projection w.r.t. that pair of layers}.  

Given a community $C \in \mathcal{C}$ and layers $L_i, L_j \in \mathcal{L}$, we will use symbols $C^{(i)}$ and $C^{(j)}$ to denote the \textit{projection} of $C$ onto the two layers, i.e., the set of nodes in $C$ that lay on $L_i$ and $L_j$, respectively. 
In the following, we define   two   approaches for measuring inter-layer couplings based on community projection. 

For any two layers $L_i, L_j$ and community $C$, the first approach, we call \textit{symmetric}, determines the relevance of inter-layer coupling of nodes belonging to $C$ as proportional to the fraction of nodes shared between $L_i$ and $L_j$ that belong to $C$. 

\vspace{2mm}
\begin{definition}
\em 
Given a community $C \in \mathcal{C}$ and layers $L_i, L_j \in \mathcal{L}$,  the \emph{symmetric projection-based inter-layer coupling}, denoted as $IC_s(C,L_i, L_j)$ and referring to term $IC$  in Eq.~(\ref{eq:MLmodularity}), is defined as the probability that $C$ lays on $L_i$ and~$L_j$:
\begin{equation}\label{eq:ICs}
\hspace{-2mm}IC_s(C,L_i, L_j)\!=\!\Pr[C \text{~in~} L_i, C \text{~in~} L_j] = \frac{|C^{(i)} \cap C^{(j)}|}{|V_i \cap V_j|}
\end{equation}
\QEDB
\end{definition}

The above definition assumes that the two events ``$C$ in $L_i$'' and  ``$C$ in $L_j$'' are independent to each other, and it   does not consider that  the coupling might have a different meaning depending on the \textit{relevance} a  community has on a particular layer in which it is located.  
By relevance of community, we simply mean here the fraction of nodes in a layer graph that belong to the community; therefore, the larger the community in a layer, the more relevant is w.r.t. that layer. 
However, we observe that \textit{more relevant community in a layer corresponds to less surprising projection from that layer to another}. This would imply that the inter-layer  coupling for that community is less interesting w.r.t. projections of smaller communities, and hence the strength of the coupling might be lowered.  
We capture the above intuition by the following definition of   \textit{asymmetric projection-based inter-layer coupling}.

 \vspace{2mm} 
 \begin{definition}
\em 
Given a community $C \in \mathcal{C}$ and layers $L_i, L_j \in \mathcal{L}$,  the \emph{asymmetric projection-based inter-layer coupling}, denoted as $IC_a(C,L_i, L_j)$ and referring to term $IC$  in Eq.~(\ref{eq:MLmodularity}), is defined as the probability that $C$ lays on $L_j$ given that $C$ lays on $L_i$:
 \begin{eqnarray}\label{eq:ICa}
IC_a(C,L_i, L_j) & = & \Pr[C \text{~in~} L_j | C \text{~in~} L_i]  =  \nonumber \\
& = & \frac{\Pr[C \text{~in~} L_i, C \text{~in~} L_j]}{\Pr[C \text{~in~} L_i]}
=  \nonumber \\
& = & \frac{|C^{(i)} \cap C^{(j)}|}{|V_i \cap V_j|} \times \frac{|V_i|}{|C^{(i)}|}
\end{eqnarray}
 \QEDB
\end{definition}


\mpara{Dealing with layer ordering. \ }
Our formulation of multilayer modularity is general enough to account for an available ordering of the layers, according to a given partial order relation. 

The previously defined asymmetric inter-layer coupling is well suited to model  situations in which we might want to express the inter-layer coupling from a ``source'' layer to a ``destination'' layer. Given any two layers $L_i, L_j$, it may be the case that only comparison of  $L_i$ to $L_j$, or vice versa, is allowed. 
This is clearly motivated when    there exist layer-coupling constraints, thus only some  of the layer couplings should  be considered in the computation of $Q$.  

In practical cases, we might assume that the layers can be naturally ordered to reflect a predetermined lexicographic order, which might be set, for instance, according to a progressive enumeration of layers or to a chronological order of the time-steps corresponding 
to the layers.  
That said,   we can consider two special cases of \textit{layer ordering}:
\begin{itemize}
\item
 \textit{Adjacent layer coupling}: 
$L_i \prec_{\mathcal{L}} L_j$ iff  $j=i+1$ according to a predetermined natural order.  
\item 
\textit{Pair-wise layer coupling}:   $L_i \prec_{\mathcal{L}} L_j$ iff  $j>i$ according to a predetermined natural order. 
\end{itemize}

\begin{figure}[t!]
\centering 
\hspace{-4.5mm}
\includegraphics[scale=0.48]{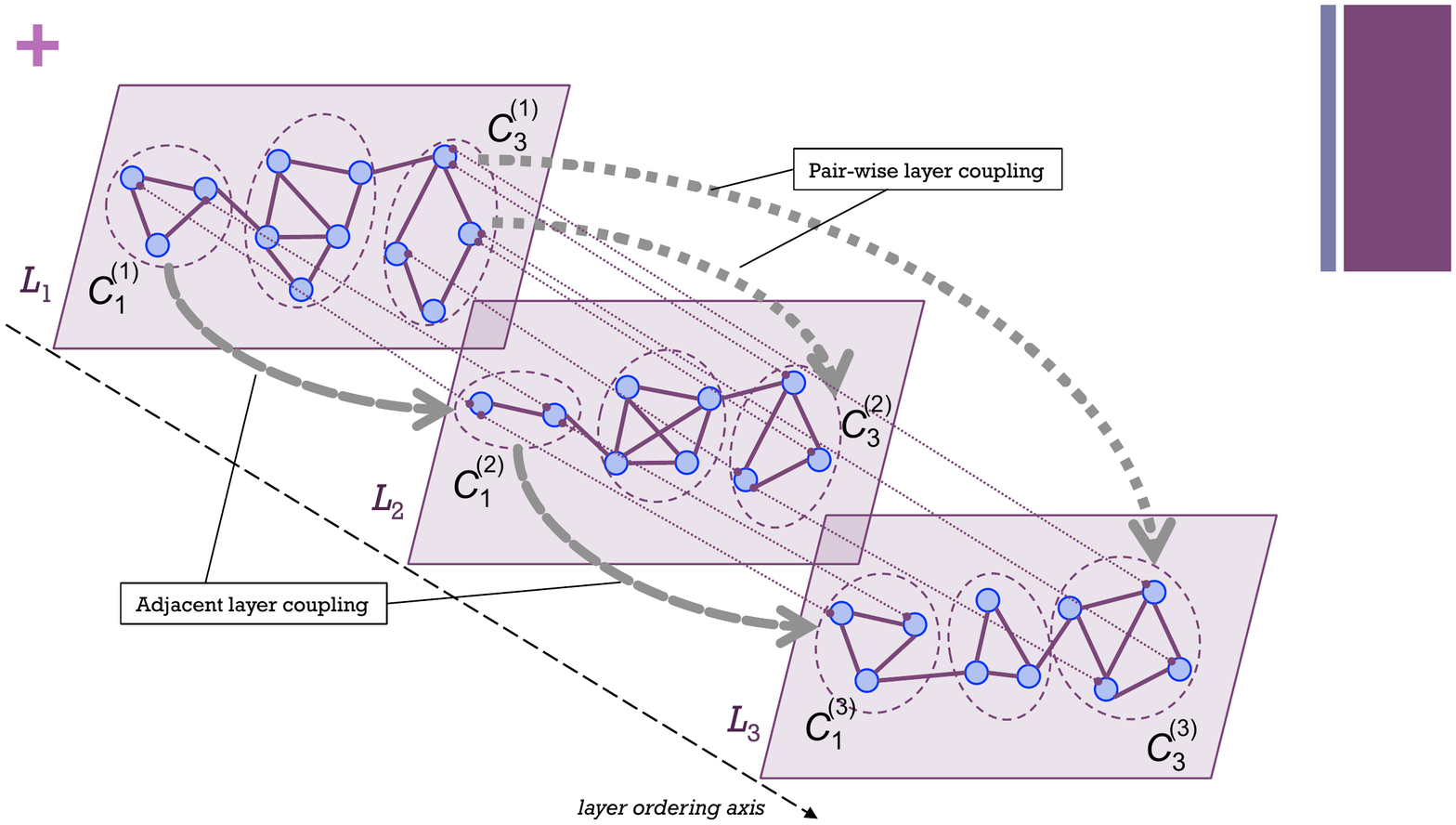}
\caption{Example multilayer network with ordered set of layers, according to   lexicographic ascendent ordering. Community $C_1$ is projected onto the three layers using adjacent layer coupling, while community $C_3$ is projected using pair-wise layer coupling. }
\label{fig:esempio}
\end{figure}

\noindent 
Note that the adjacent layer coupling scheme requires $\ell - 1$ pairs to consider, while the pair-wise layer coupling scheme involves the comparison between a layer and its subsequent ones, i.e., $(\ell^2-\ell)/2$ pairs. 
Figure~\ref{fig:esempio} illustrates an example of asymmetric inter-layer coupling over a three-layer network. 

  Moreover, it should be noted that the availability of a layer ordering enables two variants of the asymmetric projection-based inter-coupling given in Eq. (\ref{eq:ICa}).  
For any two layers $L_i, L_j \in \mathcal{L}$, such that $L_i \prec_{\mathcal{L}} L_j$ holds, we refer to as  \textbf{\emph{inner}} the direct evaluation of $IC_a(C, L_i, L_j)$, and as \textbf{\emph{outer}}  the case in which $L_i$ and $L_j$ are switched, i.e., $IC_a(C, L_j, L_i)$. 
In the inner case, the strength of coupling is higher as the projection of $C$ on the source layer (i.e., the   preceding one  in the order) is less relevant;   vice versa,  the outer case 
weights more the coupling as the projection on the destination layer  (i.e., the   subsequent one in the order) is less relevant.
For instance, considering again the example of Fig.~\ref{fig:esempio}, the   asymmetric coupling for the projection of community $C_1$ from $L_1$ to $L_2$ is stronger in the outer case, since $IC_a(C_1, L_1, L_2) = (2/9) \times (12/3)=8/9$ which is lower than  $IC_a(C_1, L_2, L_1)= (2/9) \times (9/2)=1$. 
We hereinafter use symbols $IC_{ia}$ and $IC_{oa}$ to distinguish between the inner asymmetric and the outer asymmetric evaluation cases.

\vspace{2mm}
\textit{\underline{Time-evolving multilayer networks.\ }} 
So far we have assumed that when comparing  any two layers $L_i, L_j$, with $L_i \prec_{\mathcal{L}} L_j$,  it does not matter the number of layers between $L_i$ and $L_j$. 
Intuitively, we might want to penalize the strength of the  coupling  as more ``distant'' $L_j$ is from $L_i$. This is often the case in time-sliced networks, whereby we want to understand how community structures evolve over~time. 

In light of the above remarks, we define a refinement of the asymmetric projection-based inter-layer coupling, by introducing a multiplicative factor that smoothly decreases the value of the $IC_a$ function as the temporal distance between $L_i$ and $L_j$ increases. 

\vspace{2mm}
\begin{definition}
\em 
Given a community $C \in \mathcal{C}$ and layers $L_i, L_j \in \mathcal{L}$, such that  $L_i \prec_{\mathcal{L}} L_j$, the \emph{time-aware asymmetric projection-based inter-layer coupling}, denoted as $IC_a^t(C,L_i, L_j)$, is defined as  
 \begin{equation}\label{eq:ICat}
\hspace{-2mm}IC_a^t(C,L_i, L_j)\!=\!IC_a(C,L_i, L_j) \times \frac{2}{1+ \log_2(1+j-i)}
\end{equation}
 \QEDB
\end{definition}
 Note that the second term in the above equation is 1 for the adjacent layer coupling scheme, thus making no penalization effect when only (time-)consecutive   layers are considered.

\section{Evaluation Methodology}
\label{sec:evaluation}

\subsection{Datasets}
\label{sec:datasets} 
 
We used 9 real-world multilayer network datasets. 
%
%
\textit{AUCS}~\cite{KimL15}  
describes relationships among university employees:   work together, lunch together, off-line friendship, friendship  on Facebook, and coauthorship. 
%
%
%
%
EU-Air transport  network~\cite{KimL15}  
(\textit{EU-Air}, for short) 
represents European airport connections considering  different  airlines. 
\textit{\fftwyt} (stands for FriendFeed, Twitter, and YouTube)~\cite{Magnanibook}  was built by exploiting the feature of FriendFeed as social media aggregator to align registered users who were also members of Twitter and YouTube. 
Flickr refers to the  dataset studied in~\cite{cha2009www}. 
We used the corresponding    timestamped interaction network whose links express ``who puts a favorite-marking to a photo of whom''.   
We extracted the layers on a month-basis and aggregated every six (or more)  months. 
\textit{\ghsotw} (stands for GitHub, StackOverflow and Twitter)~\cite{kdweb2015} is another cross-platform network   
where  edges express followships on Twitter and GitHub, and  ``who answers to whom'' relations on   StackOverflow.
\textit{Higgs-Twitter}~\cite{KimL15}  
represents  friendship,   reply,   mention, and retweet relations among Twitter users. 
London   transport network~\cite{ZhangWLY16}  
 (\textit{London}, for short) models three types of connections of train stations in London: underground lines, overground, and DLR.   
ObamaInIsrael2013~\cite{Omodei2015}
(\textit{Obama}, for short) models retweet, mention, and reply relations of users of  Twitter during   Obama's visit to Israel in 2013.
7thGraders~\cite{ZhangWLY16}  
(\textit{VC-Graders}, for short) represents  students involved in    friendship,  work together, and affinity relations in the class. 

Table~\ref{tab:datasets} reports for each dataset, the size of set $\V$, the number of edges in all layers, 
the average coverage of node set (i.e.,  $ 1/|\mathcal{L}| \sum_{L_i \in \mathcal{L}} (|V_i|/|\mathcal{V}|)$), and the average coverage of edge set (i.e., $1/|\mathcal{L}|  \sum_{L_i \in \mathcal{L}} (|E_i|/ \sum_{L_i} |E_i|)$). 
The table also shows basic, monoplex structural statistics (degree, average path length, and clustering coefficient)  for the layer graphs of each dataset.   
    

\begin{table*}[t!]
\caption{Main characteristics of our evaluation network datasets. Mean and standard deviation over the layers are reported for degree, average path  length, and clustering coefficient  statistics.}
\centering 
\scalebox{0.63}{
\begin{tabular}{|l||c|c|c|c|c|c|c|c|}
\hline
&\#entities &\#edges &\#layers & node set & edge set & degree & avg. path & clustering \\
& $(|\V|)$ & & $(\ell)$ & coverage & coverage &  & length & coefficient \\
\hline \hline
\textit{AUCS} & 61 & 620 & 5 &0.73 &0.20   & 10.43  $\pm$ 4.91 & 2.43 $\pm$ 0.73 &  0.43 $\pm$   0.097 \\
\hline
\textit{EU-Air} & 417 & 3\,588 & 37 & 0.13 &0.03 & 6.26 $\pm$ 2.90 & 2.25 $\pm$ 0.34 & 0.07 $\pm$ 0.08 \\
\hline
\textit{\fftwyt} & 6\,407 & 74\,836 & 3 & 0.58 &0.33 &  9.97 $\pm$ 7.27 & 4.18 $\pm$ 1.27  & 0.13  $\pm$  0.09 \\
\hline
\textit{Flickr} & 789\,019 &17\,071\,325  & 5 &0.33 &0.20 &  23.15 $\pm$   5.61 &  4.50 $\pm$    0.60 &  0.04 $\pm$   0.01  \\
\hline 
\textit{\ghsotw} &55\,140  &2\,944\,592   & 3 &0.68   &0.34   &  41.29 $\pm$ 45.09 &   3.66 $\pm$   0.62 &  0.02 $\pm$   0.01 \\
\hline
\textit{Higgs-Twitter} & 456\,631 & 16\,070\,185 & 4 &0.67  &0.25 & 18.28 $\pm$ 31.20 & 9.94 $\pm$ 9.30 & 0.003 $\pm$ 0.004\\
\hline
\textit{London} & 369 & 441 & 3 &0.36  &0.33 & 2.12 $\pm$ 0.16 & 11.89 $\pm$ 3.18 & 0.036 $\pm$ 0.032 \\
\hline
\textit{Obama} &2\,281\,259  &4\,061\,960  & 3 &0.50 &0.34 &  4.27 $\pm$ 
 1.08 &  13.22 $\pm$   4.49 &   0.001 $\pm$    0.0005  \\
\hline
\textit{VC-Graders} & 29 & 518 & 3 &1.00 &0.33 & 17.01 $\pm$ 6.85 & 1.66 $\pm$ 0.22 & 0.61 $\pm$ 0.89 \\
\hline
\end{tabular}
}
\label{tab:datasets}
\end{table*}

\subsection{Community detection methods}
\label{sec:methods}
 
We resorted to   state-of-the-art methods for  community  detection in multilayer networks, which belong to the two  major  approaches, namely  \textit{aggregation} and \textit{direct} methods. The former  detect a community structure separately for each network layer, after that  an aggregation mechanism is used to obtain the final community structure, while the latter directly work on the multilayer graph by    optimizing a multilayer quality-assessment criterion. 

As an exemplary method of the   aggregation approaches, we used  \textit{Principal Modularity Maximization} (PMM)~\cite{TangWL09}. 
 PMM aims to find a concise representation of features from the various layers (dimensions) through 
 structural feature extraction and cross-dimension integration.  
Features from each dimension are first extracted via modularity maximization, then concatenated and subject to PCA to select the top eigenvectors, which represent possible community partitions. Using these eigenvectors, a low-dimensional embedding is computed  to   capture  the principal patterns across all the dimensions of the network. The  $k$-means method is finally applied on this embedding  to discover a community structure.  
 %
As for the direct methods, we resorted to \textit{Generalized Louvain} (GL)~\cite{Mucha10} and \textit{Locally Adaptive Random Transitions} (LART)~\cite{KunchevaM15}. 
GL extends the classic Louvain method using  multislice  modularity, so that each node-layer tuple is assigned separately to a community.  Majority voting is adopted to decide the final assignment of an entity node to the  community that contains the majority of its layer-specific instances.  
LART is a   random-walk based method. It first runs a different random walk for each layer, then a dissimilarity measure between nodes is obtained leveraging the per-layer transition probabilities. A hierarchical clustering method is used   to produce a hierarchy of  communities which is eventually cut at the level corresponding to the best value of multislice modularity.

%

Note   that  PMM requires the desired number of communities ($k$) as input. 
Due to different size of our evaluation datasets, we devised several configurations of variation of parameter $k$ in PMM, by reasonably adapting each of the configuration ranges and increment steps   to the network size.  
%
%
As concerns GL and LART, we observed a tendency of LART to discover much more communities than GL, on each dataset; for instance, 381 vs. 10 on \textit{EU-Air,} 339 vs. 21 on \textit{London}, 27 vs. 5 on \textit{AUCS}. Also, the size distribution   of the communities extracted by GL is highly right-skewed (i.e., tail stretching toward the right) on the larger datasets, i.e., \textit{EU-Air}, \textit{Flickr}, \textit{Higgs-Twitter},  \textit{\fftwyt},   \textit{\ghsotw}, and \textit{Obama}, while it is moderately left-skewed on the remaining datasets.

\subsection{Experimental setting}
\label{sec:setting}

We carried out GL, PMM and LART methods on each of the network datasets and measured, for each community structure solution, our proposed multilayer modularity ($Q$) as well as the  Mucha et al.'s multislice modularity ($Q_{\textrm{ms}}$).   

 We   evaluated $Q$ using the redundancy-based resolution factor $\gamma(L,C)$ with either  the symmetric   ($IC_s$) or the asymmetric ($IC_a$) projection-based inter-layer coupling.   
%
%
 We  also considered   cases corresponding to   ordered   layers, using either the adjacent-layer scheme or the pair-wise-layer scheme, and for both schemes considering inner ($IC_{ia}$) as well as outer   ($IC_{oa}$) asymmetric coupling.  
We further evaluated the case of temporal ordering, using   the time-aware asymmetric projection-based inter-layer coupling. 
%
Yet, we considered the particular setting  of uniform resolution (i.e., $\gamma(L,C)=1$, for each layer $L$ and community $C$).  

As for $Q_{\textrm{ms}}$, we   devised two settings: the first by varying $\gamma$ within $[0..2]$ while fixing $\omega=0$ (i.e., no inter-layer couplings), the second by varying $\gamma$ and $\omega=1-\gamma$~\cite{Mucha10}.  
 

%
%

\section{Results}
\label{sec:results}


\subsection{Evaluation with unordered layers} 


 \begin{table}[t!]
\caption{Multilayer modularity $Q$ on GL community structures}\label{tab:GL}
\centering
\scalebox{0.85}{
\begin{tabular}{|c||c|c||c|c|}
\hline
  &$\gamma$, $IC_a$&$\gamma$, $IC_s$& $\gamma=1$, $IC_a$&$\gamma=1$, $IC_s$\\\hline\hline
\textit{AUCS}&    0.41 &   0.37 &      0.39  &  0.35\\\hline
\textit{EU-Air}&0.04 &   0.03 &       0.04&    0.03\\\hline
\textit{\fftwyt}&    0.50  &  0.42    &  0.42  &  0.34\\\hline
\textit{Flickr}&    0.32  &  0.31 &      0.28 &   0.27\\\hline
\textit{\ghsotw}&    0.40 &   0.40   &   0.35 &   0.35\\\hline
\textit{Higgs-Twitter}&    0.15&    0.13 &      0.14 &   0.12\\\hline
\textit{London}&    0.35  &  0.26&      0.34&    0.26\\\hline
\textit{Obama}&    0.43 &   0.32 &       0.43 &   0.32\\\hline
\textit{VC-Graders}&    0.54 &   0.53&        0.44 &   0.43\\\hline
    \end{tabular}
}
\end{table}%

Table~\ref{tab:GL} reports on values of $Q$ with different settings, measured on the community structure solutions computed by GL. 
Several remarks stand out from this table. 
With the exception of \textit{\ghsotw} on which effects on $Q$ are equivalent,  
 using $IC_a$ leads to higher  $Q$  than  $IC_s$.  
On average over all networks, using  $IC_a$ yields an increment of  13.4\% and 14.6\% (with $\gamma$ fixed to 1) w.r.t. the value of $Q$ corresponding to $IC_s$.  
 This   higher performance of $Q$ due to $IC_a$ supports our initial hypothesis on the opportunity of   asymmetric inter-layer coupling.     
It is also interesting to note that, when  fixing $\gamma$ to 1, $Q$ decreases w.r.t. our defined variable   redundancy-based resolution $\gamma(L,C)$  --- 
decrement of 11\% and 12\%   using $IC_a$ and $IC_s$, respectively.   
%

 
  \begin{table}[t!]
\caption{Multilayer modularity $Q$ on LART community structures}
\label{tab:LART}
\centering
\scalebox{0.85}{
\begin{tabular}{|c||c|c||c|c|}
\hline
 &$\gamma$, $IC_a$&$\gamma$, $IC_s$ &$\gamma=1$, $IC_a$&$\gamma=1$, $IC_s$\\\hline\hline
   \textit{AUCS}&  0.47&    0.19&         0.43 &   0.15\\\hline
\textit{EU-Air}&1.00 &   0.02&             1.00 &   0.02\\\hline
\textit{London}&    1.00&    0.01  &           1.00 &   0.01\\\hline
\textit{VC-Graders}&   0.30 &   0.28 &       0.22  &  0.20\\\hline
    \end{tabular}
}
\end{table}%
 
Table~\ref{tab:LART} shows results   from LART solutions. 
(Due to memory-resource and efficiency issues shown by the currently available implementation of LART, we are able to report  results only on some networks.\footnote{Experiments were  carried out on an Intel Core i7-3960X CPU @3.30GHz, 64GB RAM machine.})  
We observe that  the relative performance difference  between $IC_s$ and $IC_a$ settings is consistent with results found in the GL evaluation; this difference is even extreme (0.98 or 0.99) on  \textit{EU-Air} and \textit{London}, which is likely due also to the different sizes of community structures detected by the two methods (cf. Sect.~\ref{sec:methods}).  


 \begin{figure}[t!]
\centering
\begin{tabular}{ccc}
\hspace{-7mm}
\includegraphics[scale=0.1]{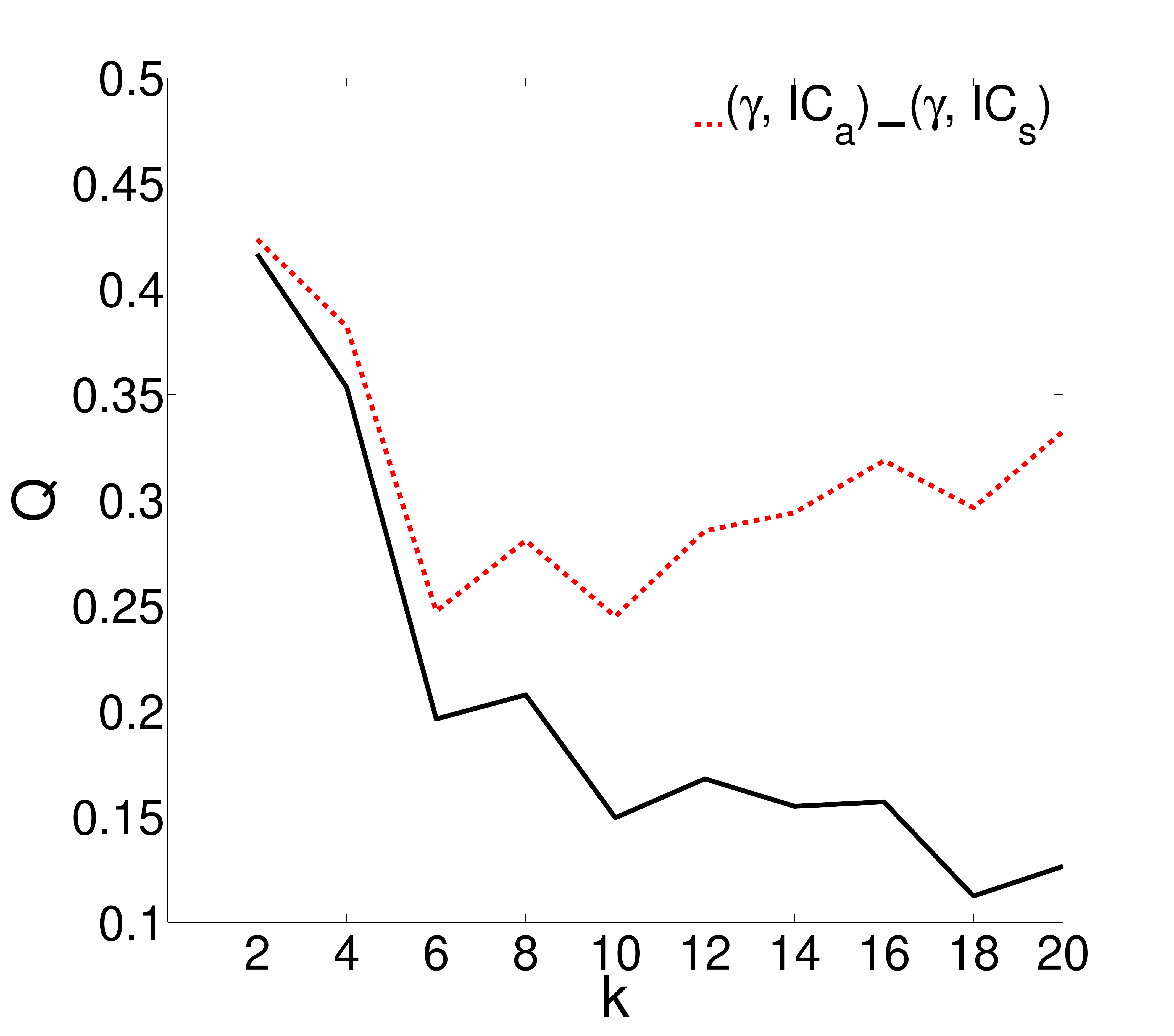} &  
\hspace{-9mm}
\includegraphics[scale=0.1]{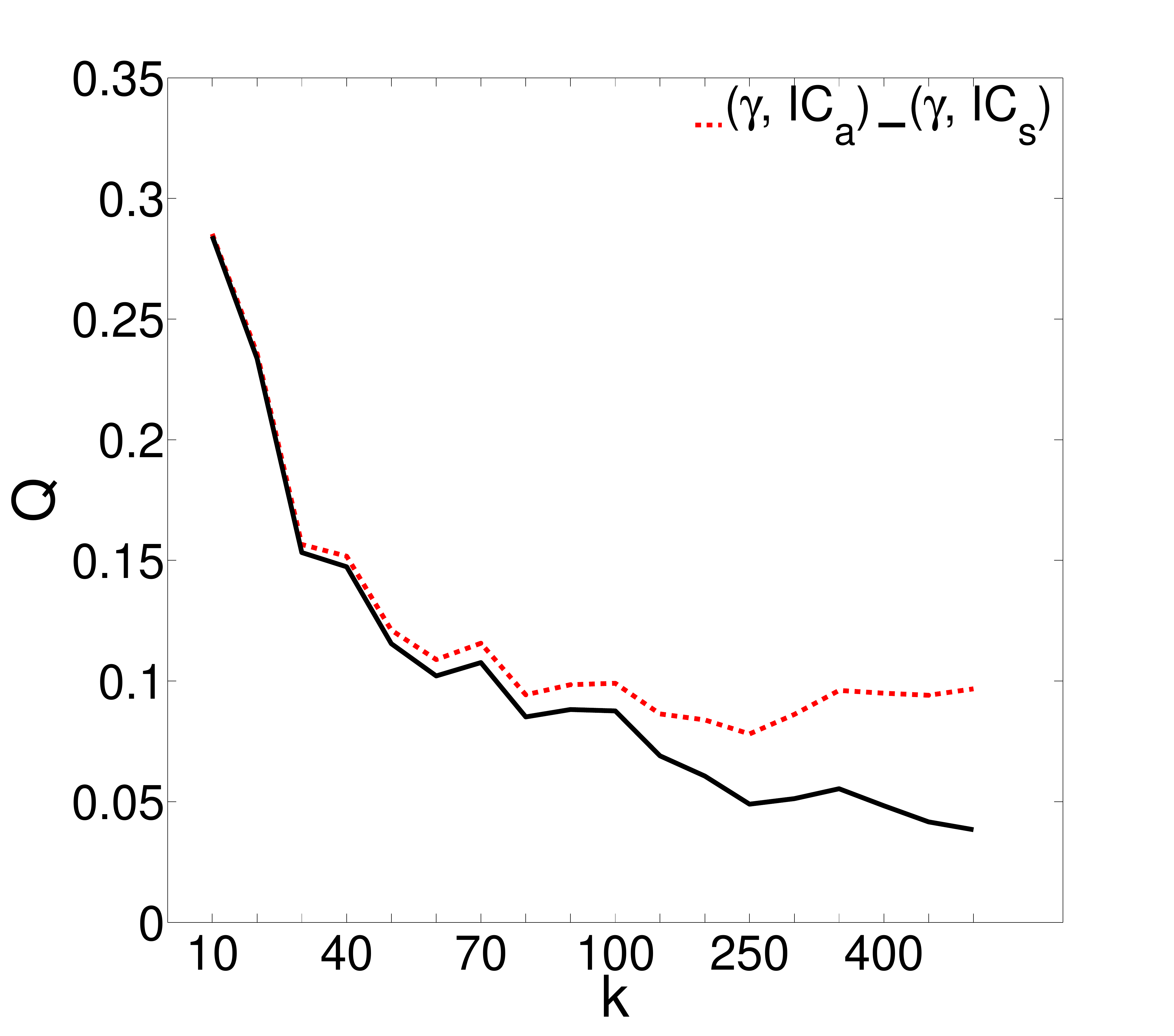} &
\hspace{-9mm}
\includegraphics[scale=0.1]{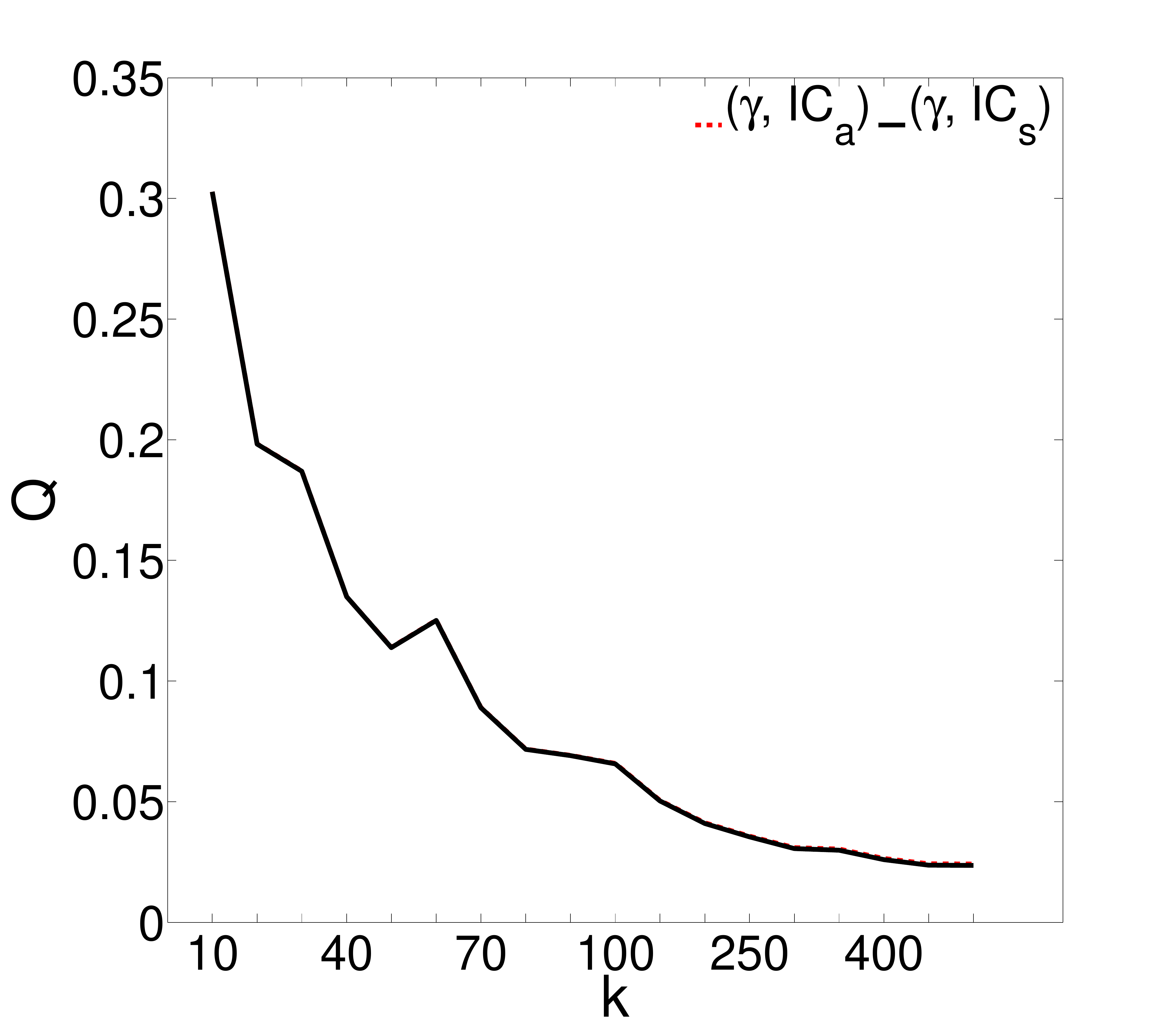} \\ 
\hspace{-7mm} (a) \textit{AUCS} & \hspace{-9mm}
 (b) \textit{\fftwyt}& \hspace{-9mm}  (c) \textit{Flickr}  \\
\hspace{-7mm}
\includegraphics[scale=0.1]{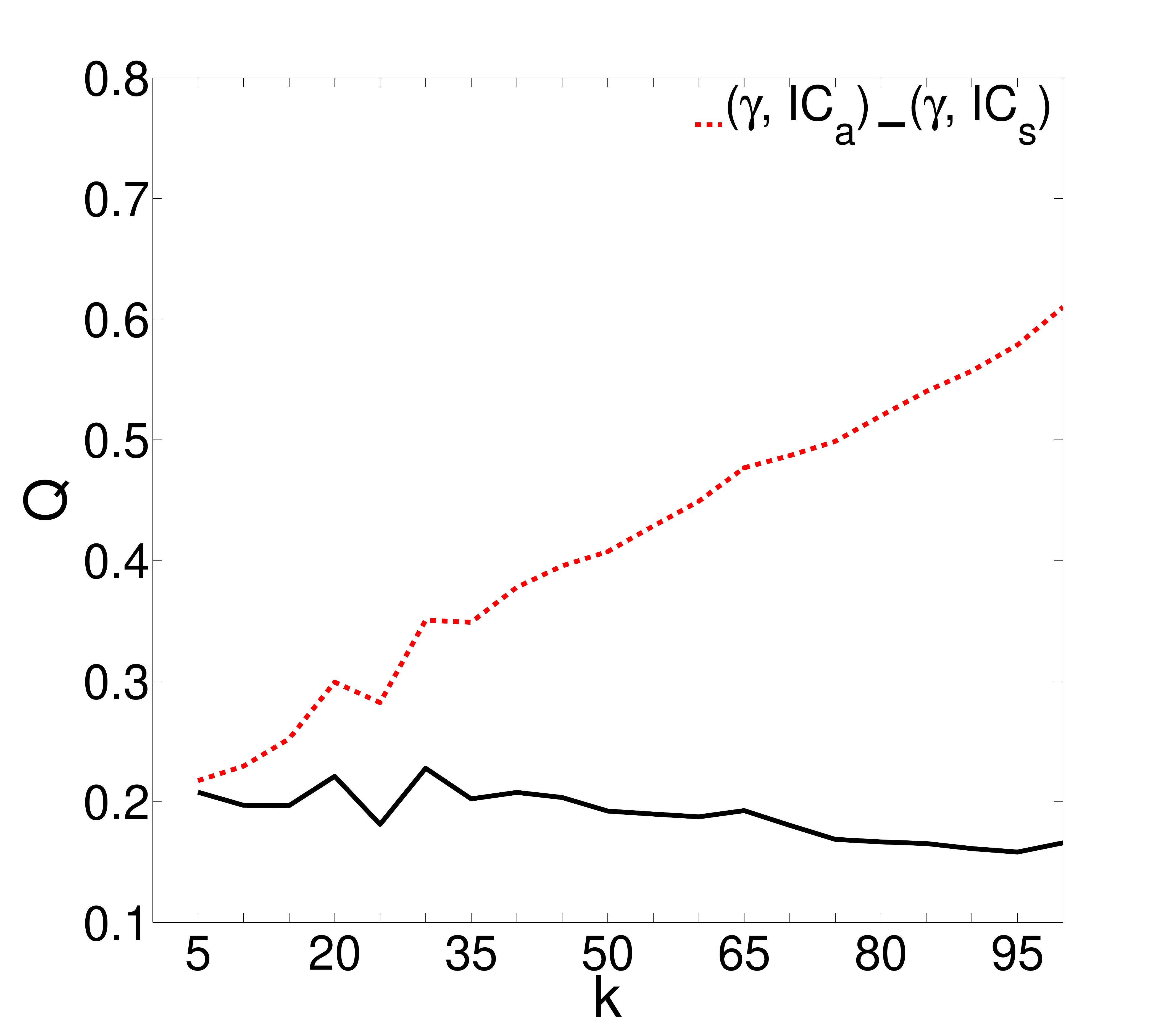} & 
\hspace{-9mm}
\includegraphics[scale=0.1]{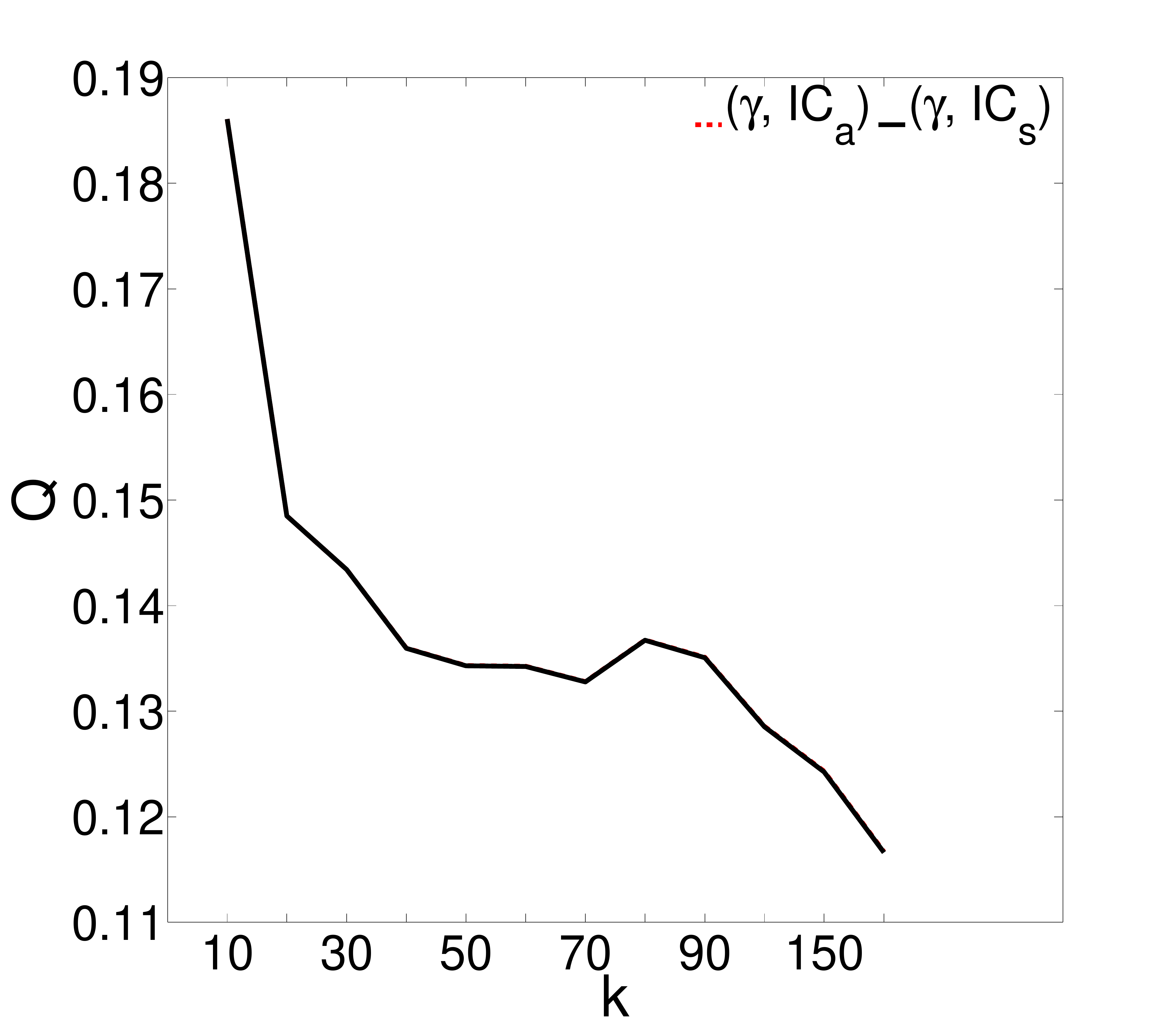} & 
\hspace{-9mm}
\includegraphics[scale=0.1]{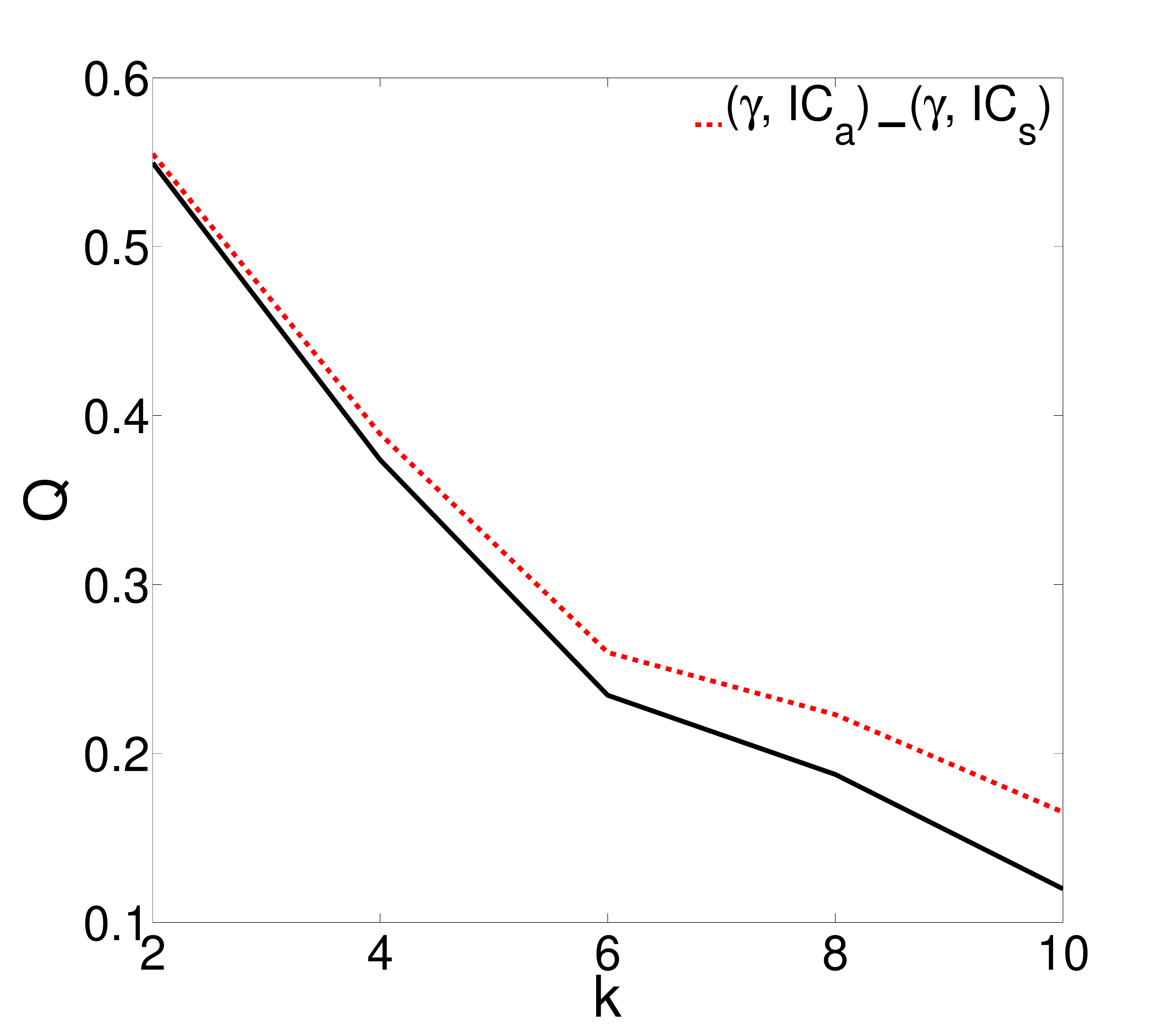} \\
\hspace{-7mm} (d) \textit{London} & \hspace{-9mm}
 (e) \textit{Obama} & \hspace{-9mm} (f) \textit{VC-Graders} \\
\end{tabular}
\caption{Multilayer modularity $Q$ on PMM community structure solutions}
\label{fig:gammaICasbeta0}
\end{figure}

 Figure~\ref{fig:gammaICasbeta0} shows how $Q$ varies in function of the number ($k$) of clusters given as input to   PMM.  
 One major remark is that  $Q$ tends to decrease as $k$ increases. This holds consistently   for the configuration of $Q$ with symmetric inter-layer coupling. 
Values of $Q$ corresponding to $IC_a$ tend to be close to the ones obtained for $IC_s$ on the largest datasets, while on the smaller ones, $IC_a$ trends  are above $IC_s$, by diverging for low $k$ in some cases; in particular,   in \textit{London} 
modularity for $IC_a$ follows a rapidly, roughly linear  increasing trend with $k$. We further inspected the behavior of $Q$ for higher regimes of $k$, which revealed that $Q$ values eventually tend to stabilize below 1.    
 As concerns the $Q$ trends over $k$ corresponding to the special setting $\gamma=1$ (results not shown), while the trends of $Q$ for $IC_a$ and for $IC_s$, respectively, do not change significantly, the values are typically lower than those obtained with redundancy-based resolution, which is again consistent with results observed for GL and LART evaluations.  
%
%
%

\subsection{Evaluation with ordered layers}


\begin{table}[t!]
\caption{Multilayer modularity $Q$, with layer ordering,  from GL and LART community structures,  on \textit{EU-Air}}
\centering
\scalebox{0.8}{
\begin{tabular}{|l||c|c|c|c||c|c|c|c|}
\hline
&\multicolumn{4}{c||}{$\gamma(L,C)$}&\multicolumn{4}{c|}{  $\gamma=1$}\\\hline
&$IC^{\textrm{Pairs}}_{ia}$&$IC^{\textrm{Pairs}}_{oa}$&$IC^{\textrm{Adj}}_{ia}$&$IC^{\textrm{Adj}}_{oa}$&$IC^{\textrm{Pairs}}_{ia}$&$IC^{\textrm{Pairs}}_{oa}$&$IC^{\textrm{Adj}}_{ia}$&$IC^{\textrm{Adj}}_{oa}$\\\hline\hline
GL  & 0.786 &0.734 &0.512  & 0.511 & 0.783&0.729&0.504&0.503 \\\hline
  LART\!\!\!& 0.981 &0.972&0.665&0.656 &0.981&0.972&0.664 &0.656 \\\hline
    \end{tabular} 
    }
\label{tab:GL-LART-Air}
\end{table}%

In this section we focus on   evaluation scenarios that  correspond to the specification of an ordering of the set of layers. 
To this purpose, we will present results on  \textit{EU-Air} and \textit{Flickr}: the former was chosen because of its highest dimensionality (i.e., number of layers) over all datasets, the latter is a time-evolving multilayer network and hence was chosen for evaluating the time-aware asymmetric inter-layer coupling. 

\begin{figure}[t!]
\centering
\begin{tabular}{cc}
\hspace{-3mm}
\includegraphics[scale=0.11]{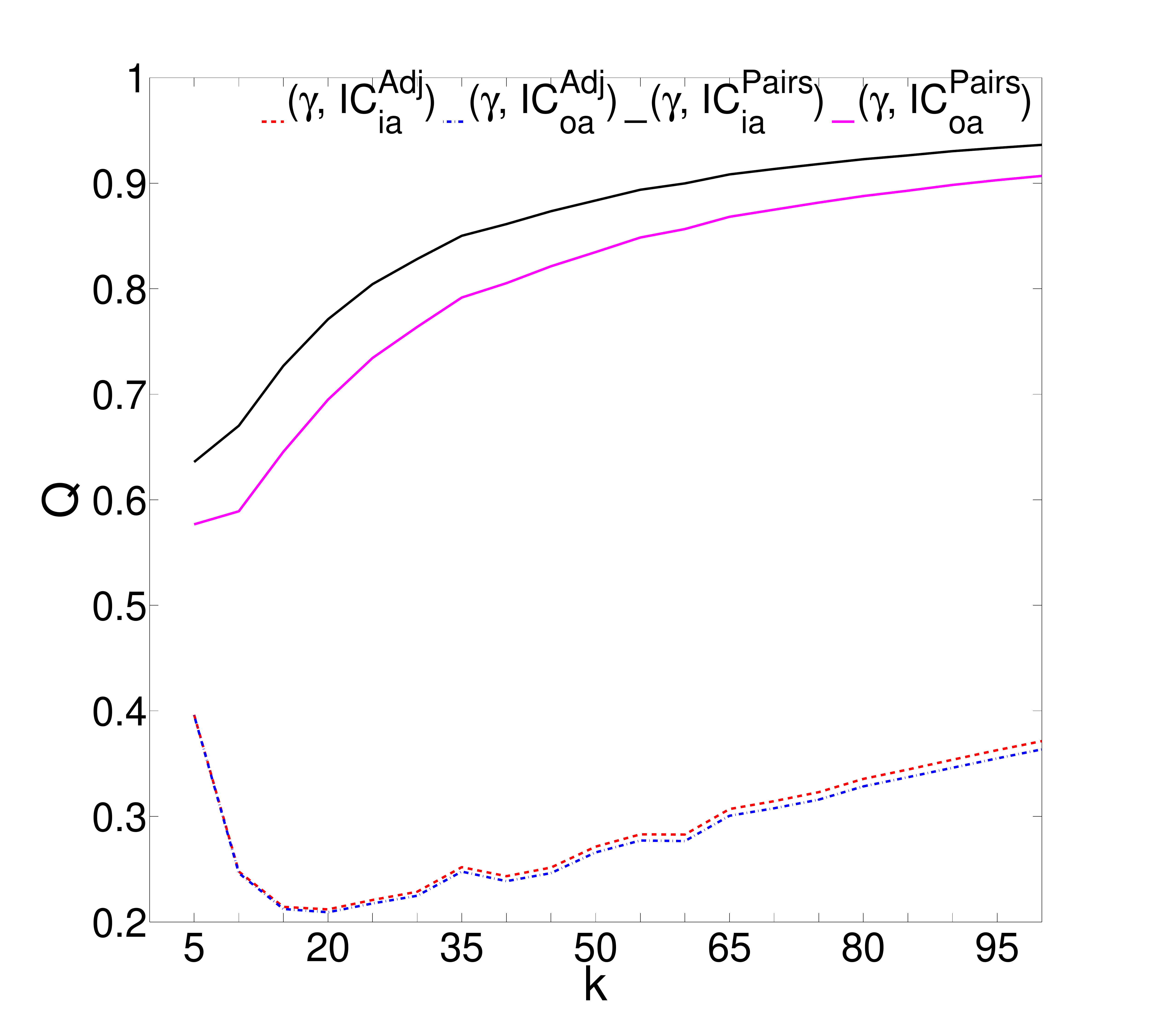} &
\hspace{-7mm}
\includegraphics[scale=0.11]{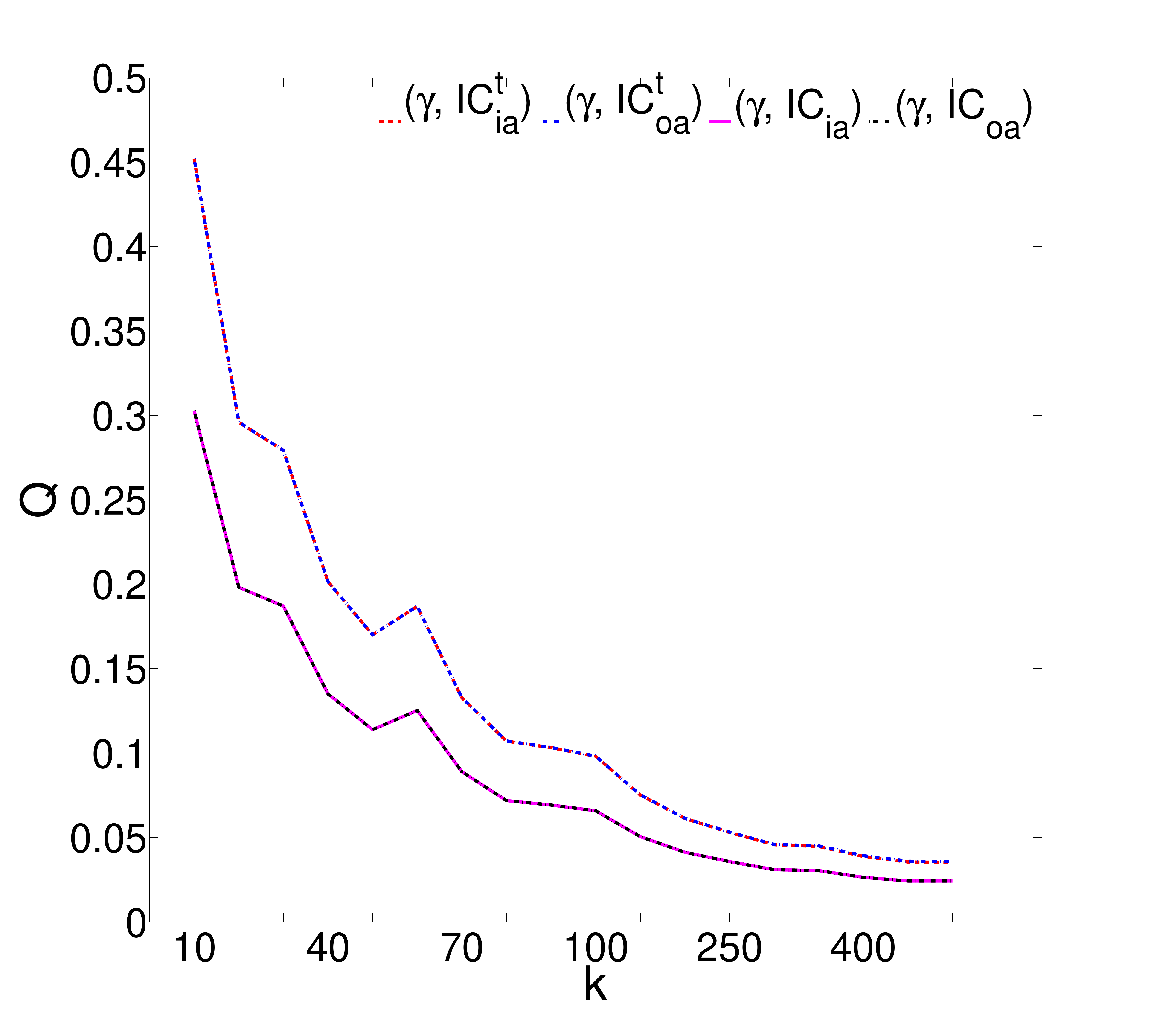} \\
\hspace{-3mm} (a) \textit{EU-Air} &  \hspace{-7mm} (b) \textit{Flickr}
\end{tabular}
\caption{Multilayer modularity $Q$ of PMM solutions   with layer ordering on (a) \textit{EU-Air} and (b)  \textit{Flickr}}
\label{fig:orderedPMM}
\end{figure}

Table~\ref{tab:GL-LART-Air} summarizes results by GL and LART on \textit{EU-Air}, corresponding to adjacent and pair-wise layer coupling.  
We observe that, in both cases of  fixed and variable resolution factor, values of $Q$ with  pair-wise layer coupling are higher than the corresponding ones for the adjacent layer coupling scheme. This would suggest that  the impact on the inter-layer coupling term is higher when all ordered pairs of layers are taken into account, than when only adjacent pairs are considered --- recall that the total degree of the multilayer graph, which normalizes the inter-layer coupling term as well, is properly computed according to the actual number of inter-layer couplings considered,  depending on whether the adjacent or the pair-wise scheme was selected. 
This result is also confirmed by PMM, as shown in Fig.~\ref{fig:orderedPMM}(a), where the plots for the pair-wise scheme superiorly bound those for the adjacent scheme, over the various $k$. 

Note also that, while the above results correspond to a descending natural ordering of the layers, by inverting this order we will have clearly  a switch between results corresponding to the inner asymmetric case with results corresponding to the outer asymmetric case.

 Figure~\ref{fig:orderedPMM}(b) compares  the effect of  asymmetric inter-layer coupling on \textit{Flickr} with and without time-awareness, for PMM solutions. We observe that  both $IC^t_{ia}$ and $IC^t_{oa}$ plots are above those corresponding to $IC_{ia}$ and $IC_{oa}$.   This indicates that considering a smoothing   term for the temporal distance between layers
   (Eq.~(\ref{eq:ICat}))   is beneficial to the increase in modularity.  
 This general result is also confirmed by GL and LART (results not shown); for instance, GL achieved on \textit{Flickr}  modularity 
 0.462 for $IC^t_{ia}$,  0.468 for $IC^t_{oa}$, and 0.460 for $IC^t_{s}$, which compared to results shown in Table~\ref{tab:GL}  represent increments in $Q$ of    43\%.

\subsection{Analysis of $Q_{\textrm{ms}}$ and qualitative comparison with $Q$} 
 We discuss here performance results obtained by the community detection algorithms with   $Q_{\textrm{ms}}$ as assessment criterion. We will refer to the default setting of unordered set of layers as stated in~\cite{Mucha10}. 
  
\begin{figure}[t!]
\centering
\begin{tabular}{cc}
\hspace{-7mm}
\includegraphics[scale=0.13]{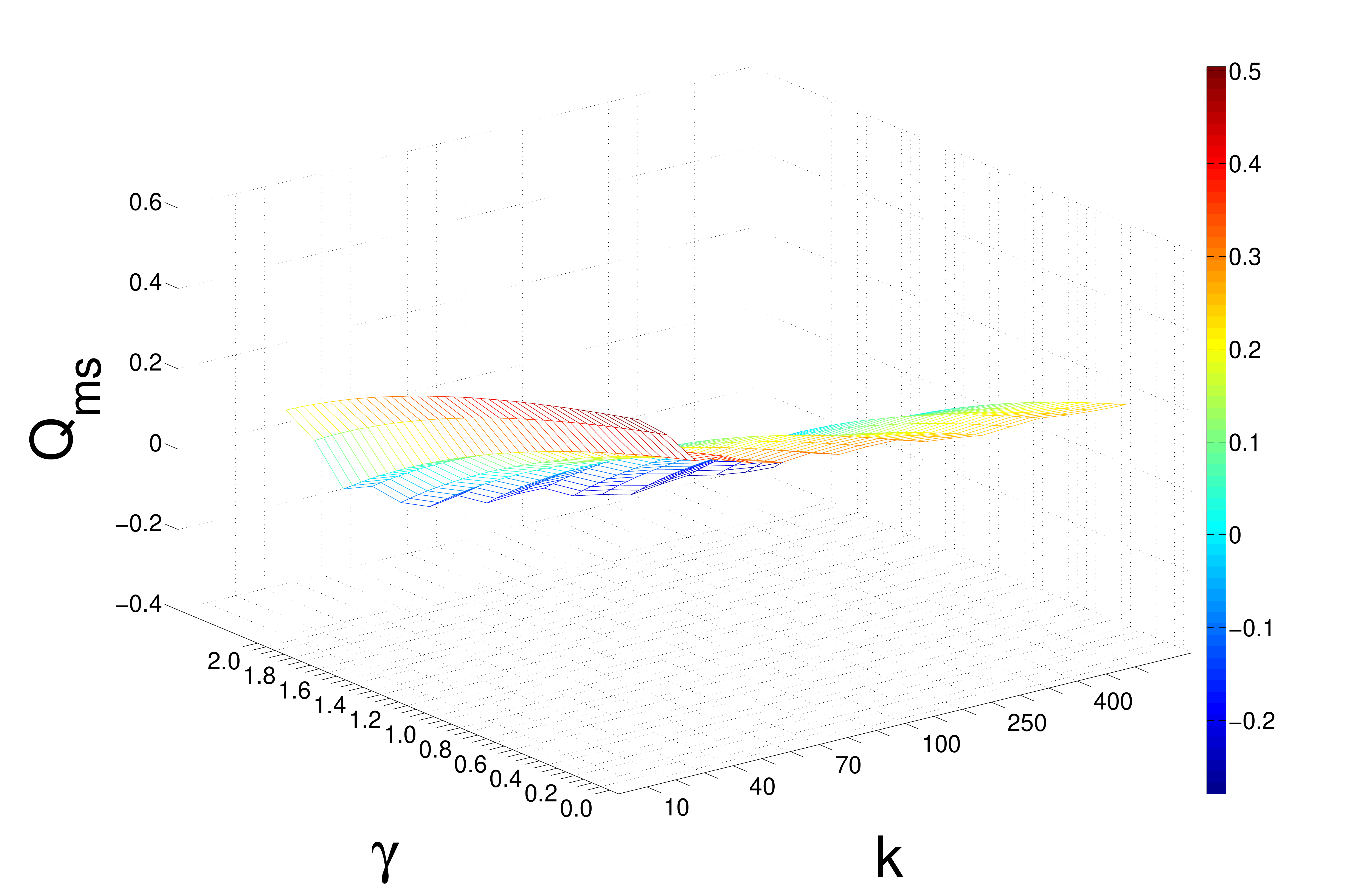} & 
\hspace{-7mm}
\includegraphics[scale=0.13]{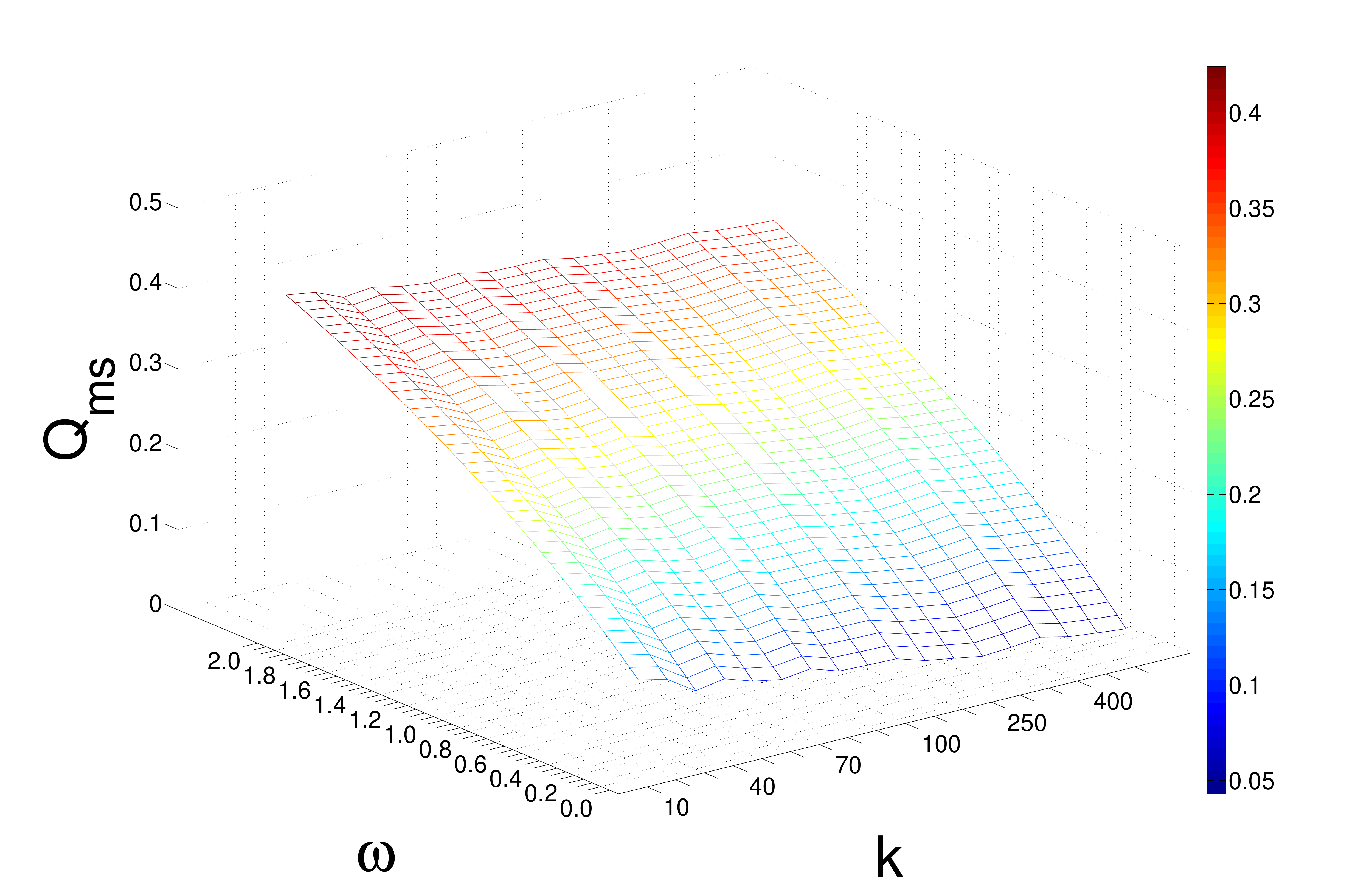} \\
(a) \textit{\fftwyt} & (b) \textit{\fftwyt}\\
\hspace{-7mm}\includegraphics[scale=0.13]{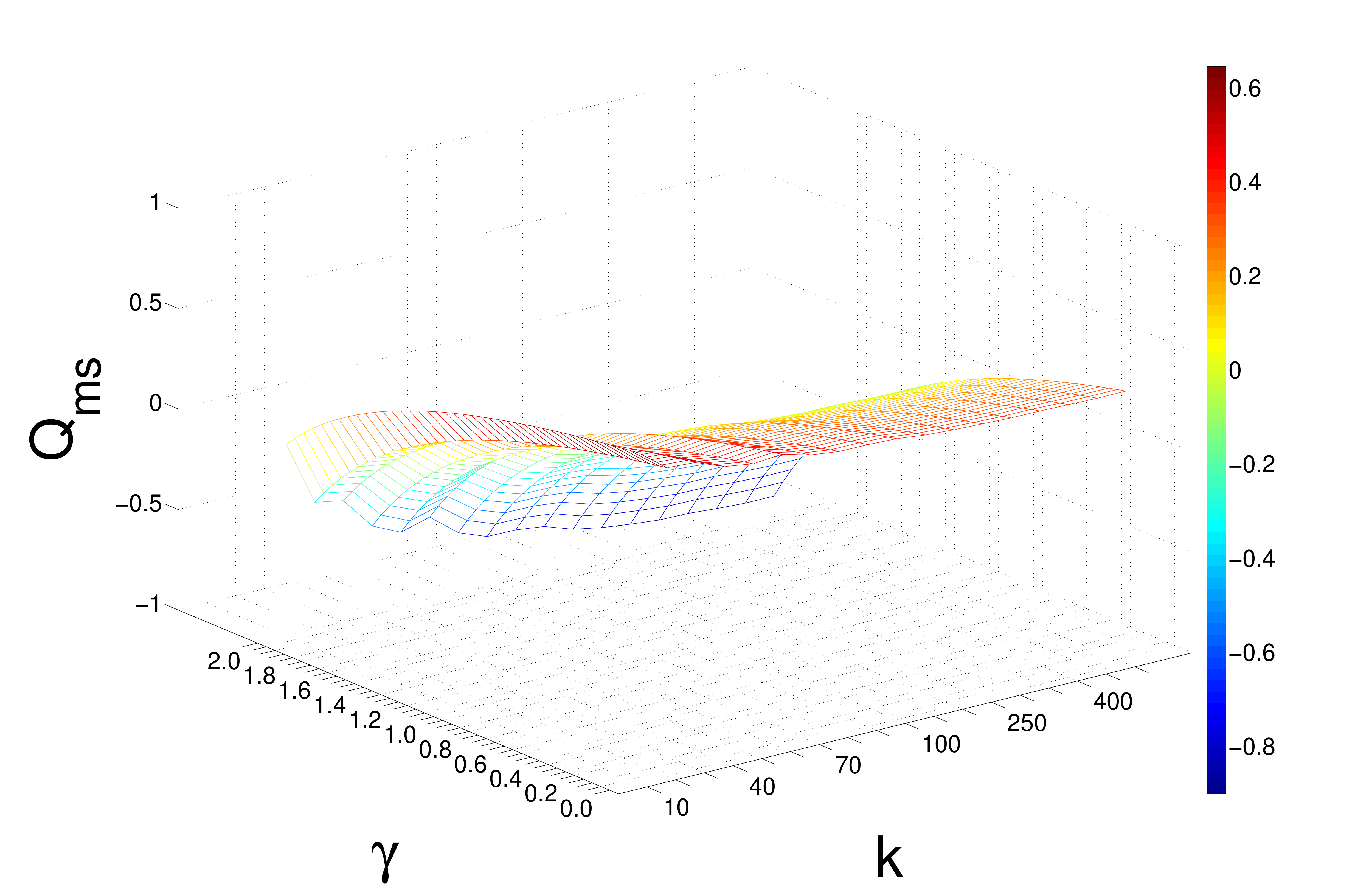} & 
\hspace{-7mm}
\includegraphics[scale=0.13]{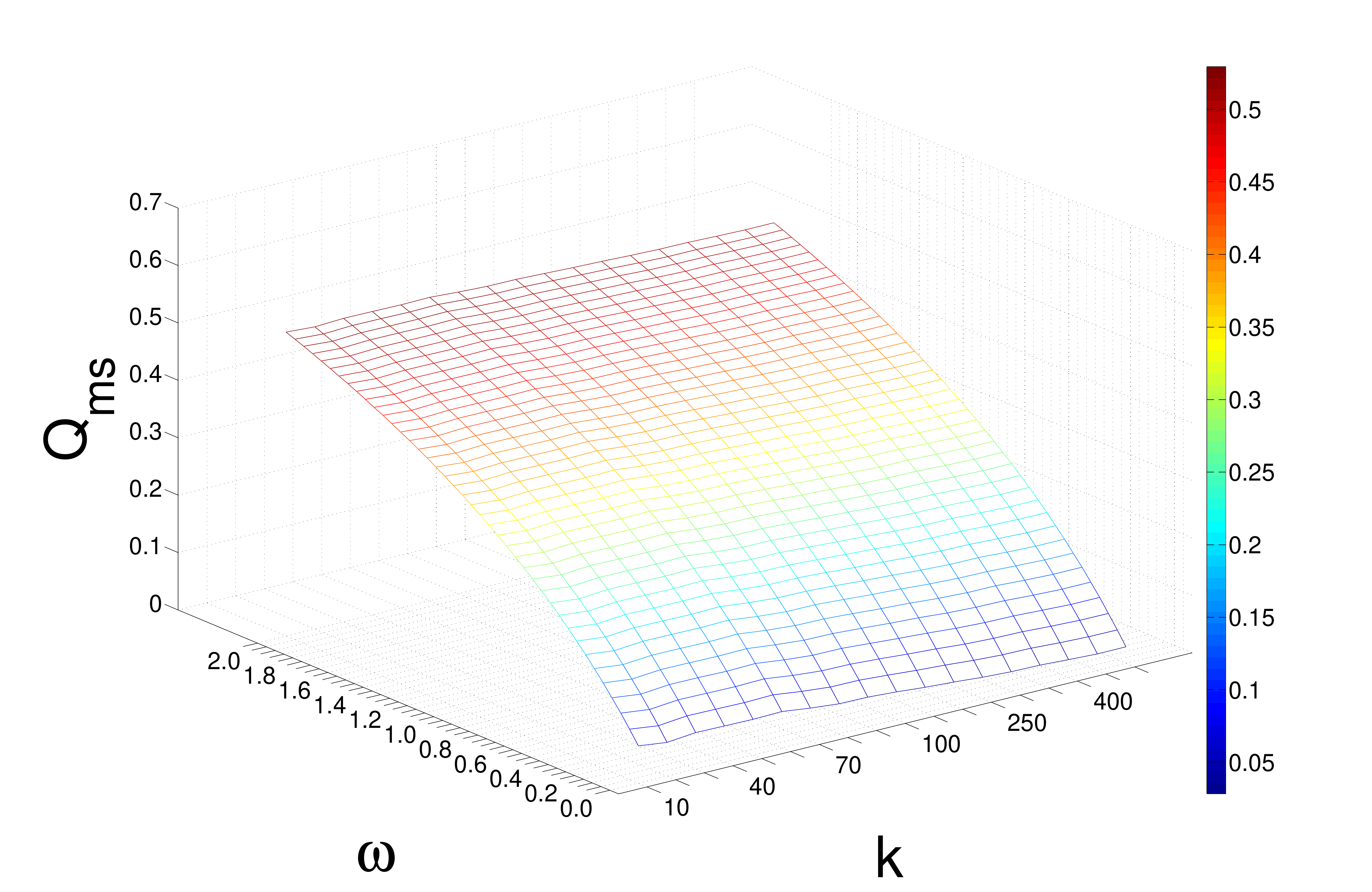} \\
(c) \textit{Flickr} & (d) \textit{Flickr} \\
\end{tabular}
\caption{Performance of Mucha et al.'s   modularity ($Q_{\textrm{ms}}$) by varying $\gamma$ with $\omega=1-\gamma$ (left) and by varying $\omega$ with $\gamma=1$ (right)}
\label{fig:Mucha}
\end{figure}

Using GL,  $Q_{\textrm{ms}}$ tends to decrease as $\gamma$ increases (while $\omega$ decreases, as it was varied with $\gamma$ as $\omega=1-\gamma$). This occurs monotonically in most datasets, within positive ranges (e.g.,    from 0.636 to 0.384 on \textit{\fftwyt},  from 0.525 to 0.391 on \textit{\ghsotw}) or including negative modularity for higher $\gamma$ (e.g., 
from 0.645 to -0.05 on \textit{Flickr}, from 0.854 to -4 on \textit{AUCS}). 
Remarkably, the simultaneous effect of  $\gamma$ and  $\omega=1-\gamma$  on  $Q_{\textrm{ms}}$ leads on some datasets (\textit{Obama}, \textit{EU-Air}, \textit{London}) to  a drastic degradation of modularity (down to much negative values) for some     $\gamma >1$, followed by a rapid increase to modularity of 1 as $\gamma$ increases closely to~2.   
Analogous considerations hold for LART and PMM. For the latter method, the plots on the left side of Fig.~\ref{fig:Mucha} show results by varying $k$, from a selection of datasets.
  Surprisingly, it appears that $Q_{\textrm{ms}}$ is  relatively less sensitive to the variation in the community structure than our $Q$. 

As shown in the plots on the right side of Fig.~\ref{fig:Mucha}, when varying $\omega$ within [0..2], with $\gamma=1$, $Q_{\textrm{ms}}$ tends to monotonically increase as $\omega$ increases. This holds   consistently on all datasets and for all methods. Variations are always on positive intervals (e.g.,   from 0.248 to 0.621 on \textit{Flickr}, from 0.305 to 0.541 on \textit{\fftwyt}, from 0.136 to 0.356 on \textit{Higgs-Twitter}). 
  
It should be noted that, while we could not directly compare the values of $Q$ and $Q_{\textrm{ms}}$, a few interesting remarks arise by observing their different behavior over the same  community structure solutions, in function of the resolution and inter-layer coupling factors. From a qualitative viewpoint, the effect  of $\omega$ on  $Q_{\textrm{ms}}$ turns out to be opposite, in most cases, to the effect of our $IC$ terms on $Q$: that is, accounting more for the inter-layer couplings leads to increase    $Q_{\textrm{ms}}$, while this does not necessarily happen in $Q$. 
Less straightforward is comparing the use of a constant resolution for all layers, as done in $Q_{\textrm{ms}}$, and the use of variable (i.e., for each pair of layer and community)  resolutions, as done in $Q$. In this regard, we have previously observed that the use of a varying redundancy-based resolution factor improves $Q$ w.r.t. the setting $\gamma=1$. By coupling this general remark with the results (not shown) of an inspection of the values of $\gamma(L,C)$   in the computation of $Q$ on the various network datasets (which confirmed   that $\gamma(L,C)$ values span over its range,  in practice), we can conclude that a more appropriate   consideration of the term modeling the expected connectivity of community is realized in our $Q$ w.r.t. keeping the resolution as constant for all layers in $Q_{\textrm{ms}}$.

%
%
%
%
%
%
%
%
%


 \vspace{2mm}
\section{Conclusion}

We proposed a new definition of modularity for multilayer networks. Motivated by the necessity of  a revision of the existing multislice modularity proposed in~\cite{Mucha10},  
 we conceive and formally define new notions of layer resolution and inter-layer coupling, which are essential to a generalization of modularity for multilayer networks.  
Using 3 state-of-the-art methods for multilayer community detection and 9 real-world multilayer networks, we provided empirical evidence of the significance of our proposed modularity. 
Our work paves the way for the development of new optimization methods of community detection in multilayer networks, which by embedding our multilayer modularity, can discover     community structures having the interesting   properties relating to the proposed per-layer/community redundancy-based resolution factors and projection-based inter-layer coupling schemes.




 \vspace{3mm}

%

\end{document}